\documentclass[longbibliography, aps, pra, amsmath, amssymb, amsfonts, twocolumn,superscriptaddress, footinbib]{revtex4-2}
\usepackage[german,english]{babel}
\usepackage{graphicx}
\usepackage{graphics}
\usepackage{dcolumn}
\usepackage{bm}
\usepackage{amssymb}
\usepackage{amsmath}
\usepackage{amsfonts}
\usepackage{epsfig}
\usepackage{mathbbol}
\usepackage{latexsym}
\usepackage{dcolumn}
\usepackage{subfigure}
\usepackage{hyperref}
\usepackage{float}
\usepackage[normalem]{ulem}
\usepackage[usenames, dvipsnames]{color}
\usepackage{epstopdf}
\usepackage{tabularx}
\usepackage{youngtab}
\usepackage{subfigure}
\usepackage[ruled]{algorithm2e}

\usepackage{theorem}

\hypersetup{                    
    colorlinks,
    citecolor=blue,
    filecolor=blue,
    linkcolor=blue,
    urlcolor=blue
}
\usepackage[utf8]{inputenc}

\begin{document}
\title{Mitigating algorithmic errors in quantum optimization through energy extrapolation}

\author{Chenfeng Cao}
\email{chenfeng.cao@connect.ust.hk}
\affiliation{Department of Physics, The Hong Kong University of Science and Technology, Clear Water Bay, Kowloon, Hong Kong, China}

\author{Yunlong Yu}
\affiliation{ State Key Laboratory of Low Dimensional Quantum Physics, Department of Physics,\\Tsinghua University, Beijing 100084, China}
\affiliation{Kavli Institute for Theoretical Sciences, University of Chinese Academy of Sciences, Beijing 100190, China}

\author{Zipeng Wu}%
\affiliation{Department of Physics, The Hong Kong University of Science and Technology, Clear Water Bay, Kowloon, Hong Kong, China}

\author{Nic Shannon}
\affiliation{Theory of Quantum Matter Unit, 
	Okinawa Institute of Science and Technology Graduate University, Onna-son, Okinawa 904-0412, Japan}

\author{Bei Zeng}%
\email{zengb@ust.hk}
\affiliation{Department of Physics, The Hong Kong University of Science and Technology, Clear Water Bay, Kowloon, Hong Kong, China}

\author{Robert Joynt}
\email{rjjoynt@wisc.edu}
\affiliation{Department of Physics, University of Wisconsin–Madison, 1150 University Avenue, Madison, Wisconsin 53706, USA}
\affiliation{Kavli Institute for Theoretical Sciences, University of Chinese Academy of Sciences, Beijing 100190, China}

\begin{abstract}
Quantum optimization algorithms offer a promising route to finding the ground states of target 
Hamiltonians on near-term quantum devices. 
Nonetheless, it remains necessary to limit the evolution time and circuit depth as much 
as possible, since otherwise decoherence will degrade the computation. 
Even when this is done, there always exists a non-negligible error in estimates of the 
ground state energy. 
Here we present a scalable extrapolation approach to mitigating this algorithmic error, which 
significantly improves estimates obtained using three well-studied quantum optimization 
algorithms: quantum annealing (QA), the variational quantum eigensolver (VQE), and the quantum imaginary time evolution (QITE) at fixed evolution time or circuit depth.
The approach is based on extrapolating the annealing time to infinity or the variance of estimates to zero. 
The method is reasonably robust against noise, and for Hamiltonians which only involve few-body interactions, the additional computational overhead is an increase in the number of measurements by a constant factor. 
Analytic derivations are provided for the quadratic convergence of estimates of energy as a function of time in QA, and the linear convergence of estimates as a function of variance in all three algorithms. 
We have verified the validity of these approaches through both numerical simulation and experiments on IBM quantum machines. 
This work suggests a promising new way to enhance near-term quantum computing through classical post-processing.
\end{abstract}
\date{\today}
\maketitle

\section{Introduction}

Obtaining a precise estimate of the ground state energy of a many-body Hamiltonian is a central task in modern physics and chemistry~\cite{mcardle2020quantum}. 
Classical algorithms have had considerable success by using, for example, tensor-network methods, 
but for large systems with generic long-range entanglement, the task remains challenging. 
There is hope that eventually, fault-tolerant quantum computers will be able to deal with such problems more accurately and efficiently. 
But for the time being, we remain in the era of \textit{noisy intermediate-scale quantum} (NISQ) 
computation~\cite{preskill2018quantum, deutsch2020harnessing}. 
A number of quantum algorithms have been developed which can be implemented on existing 
NISQ processors, including \textit{quantum annealing} (QA)~\cite{finnila1994quantum, kadowaki1998quantum, albash2018adiabatic}, 
the \textit{variational quantum eigensolver} (VQE)~\cite{peruzzo2014variational, kandala2017hardware, parrish2019quantum}, 
and the \textit{quantum imaginary time evolution} (QITE)~\cite{motta2020determining,mcardle2019variational,sun2021quantum, cao2021quantum}. 
Each method shows promise but is limited by the resources available to support them.


QA is a heuristic approach to finding the ground state of a complex quantum system, inspired by the adiabatic theorem. 
In QA, the ground state of a simple Hamiltonian $H_{\mathrm{init}}$, is slowly evolved into
the ground state of the target Hamiltonian $H_f$. 
The time taken to accomplish this ``annealing'' grows rapidly with the inverse of the minimum spectral gap encountered during the evolution and may become very long if this gap is small.
This can render accurate QA calculations very difficult on NISQ processors. 


Another class of NISQ algorithms that has attracted particular attention in recent years are variational quantum algorithms~\cite{cerezo2021variational,bharti2021noisy,peruzzo2014variational, kandala2017hardware, parrish2019quantum,romero2017quantum,cao2021noise,zeng2021variational,farhi2014quantum,yu2021quantum,cerezo2020variational,chen2020variational, cao2022quantum}. 
Unlike QA, these algorithms also require a classical computer to carry out an 
optimization that complements the quantum processing. 
VQE is the prime example of a variational approach to the ground--state preparation. 
In VQE, an initial product state is prepared and then evolved using a variational quantum circuit, whose parameters 
are updated iteratively, so as to minimize the expectation value of the Hamiltonian.
Where this process is successful, the result is an estimate of the ground state energy $E_{\mathrm{gs}}$.


VQE has achieved great success in electronic structure calculations~\cite{arute2020hartree}. 
However, its applications in large systems are severely hampered by three daunting obstacles.   
The first of these is the \textit{barren plateau} phenomenon: for a deep random quantum circuit, 
the gradient of the cost function used in classical minimization vanishes exponentially 
with increasing system size ~\cite{mcclean2018barren, cerezo2021cost}. 
The second is the hardware noise, which can accumulate exponentially with increasing circuit depth and induce 
another kind of ``barren plateau"~\cite{wang2020noise}. 
Quantum error-mitigation techniques have been developed to mitigate the effect of 
hardware noise~\cite{li2017efficient, endo2018practical, temme2017error, kandala2019error, koczor2021exponential, PhysRevX.11.041036}.
However, their overhead scales exponentially with circuit depth~\cite{takagi2021fundamental}. 
Consequently, VQE is only reliable for shallow quantum circuits with limited expressive power.  
Schemes to enhance shallow-depth VQE are therefore highly desirable. The third obstacle is the local minima problem~\cite{anschuetz2022beyond}. VQE for a wide class of models is swamped with local traps. Without a good guess, even if the variational circuit is capable of preparing the exact ground state, we need to sample a mass of initial parameters to prepare the ground state and estimate the ground state energy accurately.


In the QITE algorithm, ground states are prepared by using a quantum computer to simulate the process of imaginary--time evolution~\cite{motta2020determining, mcardle2019variational}. 
The evolution of a state in imaginary time is split into many steps 
via the Trotter-Suzuki decomposition, with each non-unitary time--step being 
approximated by a unitary evolution. 
Besides the ground state energy, QITE has also been used to calculate complex 
observables such as correlation functions~\cite{sun2021quantum}. 
Unfortunately, QITE is even less robust against noise than VQE, since 
errors that accumulate in earlier steps severely affect the accuracy 
of later ones. 
For this reason, one can only reliably execute QITE for very small numbers 
of Trotter steps.  In addition, the number of measurements required to perform QITE does not scale favorably with system size.


\begin{figure}[t]
	\centering
	\includegraphics[width=7.5cm]{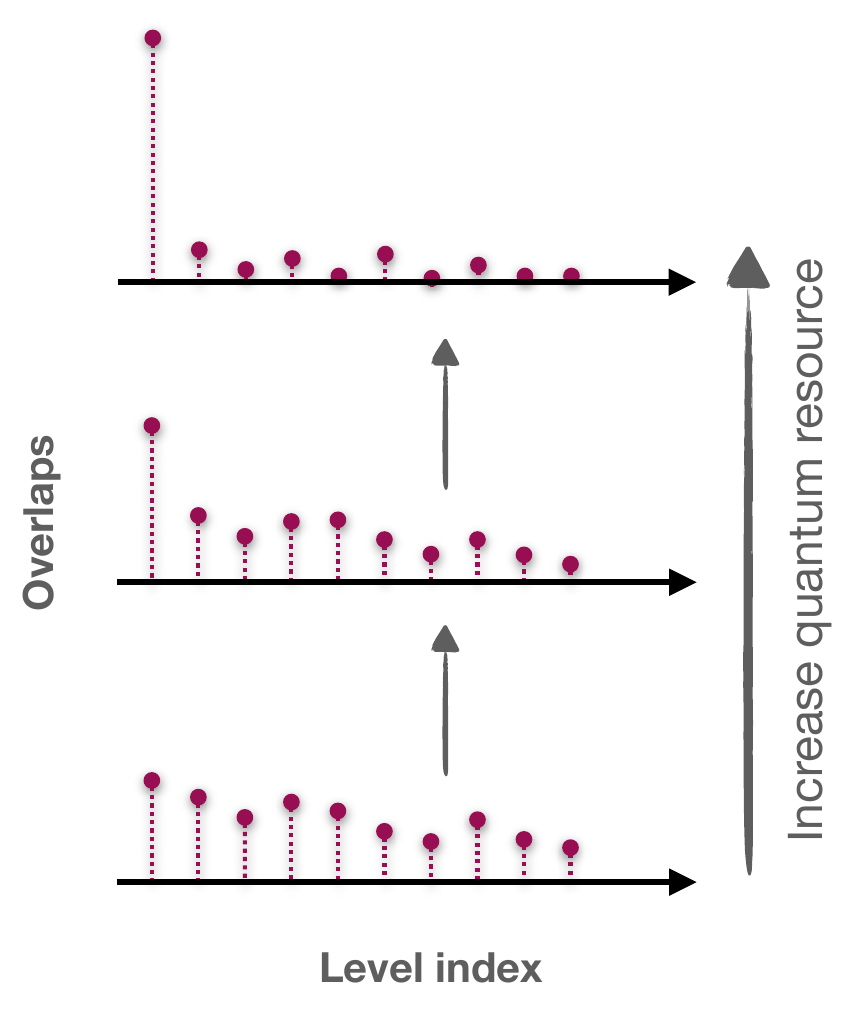}
	\caption{Depiction of state probabilities on an energy level diagram for different quantum resources (evolution time, quantum circuit depth). With the increase of resources, the state obtained in the calculation has larger overlap with the ground state and smaller overlaps with the excited states.}
	\label{fig:schematic}
\end{figure}


One question that naturally arises is whether we can infer $E_{\mathrm{gs}}$ 
from shorter evolution times in QA, or lower circuit depth for VQE/QITE? 
We take a clue from a general picture of the state probabilities on an 
energy-level diagram as the quantum resources increase.  
This is shown in Fig.~\ref{fig:schematic}. 
With the increase of resources, the state obtained in the calculation has 
a larger overlap with the ground state and smaller overlaps with the excited states. 
Accordingly, the energy converges to $E_{\mathrm{gs}}$, and the variance of the energy 
converges to 0. 
This pattern suggests two possible extrapolations to the limit of infinite 
resources when the probabilities condense entirely to the ground state.  
In the case of QA, we can estimate the ground state energy in the limit 
of infinitely-slow annealing, by carrying out estimates for different 
annealing times, and then extrapolating results to the limit of 
infinite time.
We will call this the \textit{(infinite-)time method}.  
Alternatively, for quantum optimization algorithms, we can measure both energy and variance over repeated trials, and then extrapolate results for energy as a function of variance, to obtain an estimate of the ground-state energy 
in the limit of zero variance.
We will refer to this as the \textit{(zero-)variance method}. 


Extrapolation techniques have a long history in computational and nuclear
physics~\cite{shimizu2010novel, shimizu2012variational, iqbal2013gapless}, 
and zero-variance extrapolation is a well-established approach in quantum Monte Carlo
simulation~\cite{kwon1998effects, sorella2001generalized}. 
In the context of quantum computing, zero-noise extrapolation schemes have also 
been developed to mitigate the effect of hardware 
noise~\cite{li2017efficient, endo2018practical, he2020zero}, 
and Trotter error~\cite{endo2019mitigating}. 
We note that zero-variance extrapolation has also recently been independently proposed in the context of VQE~\cite{kreshchuk2021quantum}.
However, that work did not address the role of noise or 
circuit depth in zero-variance extrapolation, or provide an
explicit experimental demonstration of its effectiveness.
Moreover, extrapolations as a function of time remain largely 
unexplored in the context of quantum computation.


The goal of the present article is to establish the zero-variance and infinite-time extrapolation methods as a practical resource 
for quantum optimization on NISQ devices.
We introduce each method in turn, discussing the expected asymptotic 
behavior, the role of circuit depth, and the consequences of noise.
The effectiveness of each method is explored through explicit 
numerical simulation.
For the zero-variance method, we also demonstrate 
its application in experiments on IBM quantum computers.



We find that both zero-variance and infinite-time extrapolations 
can effectively mitigate algorithmic errors induced by shallow quantum 
circuit depth (short decoherence time). 
%
We also find that both methods are robust against the effects 
of measurement errors. 
We further show that the zero-variance extrapolation is reasonably robust against 
gate errors, 
and that the inclusion of small but finite gate errors can even lead to 
improved estimates of ground state energy. 
%
%
We conclude that both time and variance extrapolation methods show great potential
for application to quantum optimization problems in the NISQ era.


The paper is structured as follows. In Sec.~\ref{Sec:methods}, we introduce our methods and analyze the noise effect. In Secs.~\ref{Sec:EE_QA},~\ref{Sec:EE_VQE},~\ref{Sec:EE_QITE}, we numerically and experimentally demonstrate the accuracy improvement in QA, VQE, and QITE. The conclusions and future directions are summarized and discussed in Sec.~\ref{Sec:Conclusions}.  An analytic explanation for the quadratic convergence feature of the residual energy in QA is given in Appendix~\ref{Sec:Appendix1}. An analytic derivation for the linear convergence of the energy variance is given in Appendix~\ref{Sec:Appendix2}.

\section{Methods}\label{Sec:methods}
\subsection{Time Method}
In the QA framework, we start from the ground state of a simple initial Hamiltonian $H_{\mathrm{init}}$ and evolve the system according to the time-dependent Hamiltonian 
\begin{equation}
    H(t) = f(t)H_{\mathrm{init}} + g(t)H_f,
\end{equation}
where $H_f$ is the problem Hamiltonian.  $f(t)$ and $g(t)$ are continuous functions that satisfy $f(0)=1,f(t_a)=0,g(0)=0,g(t_a)=1$, where $t_a$ is the final time. According to the adiabatic theorem, if there is no symmetry-related level crossing between the ground state and the first excited state during the evolution and the annealing time is long enough, we can exactly prepare the ground state of $H_f$.  

An actual measurement at finite $t_a$ gives the measured energy $E = \langle \psi_f|H_f|\psi_f \rangle$ in the final state $|\psi_f \rangle$.  The total probability of finding the system in an excited state of $H_f$ in the measurement is of order $\mathcal{O}(1/t_a^2)$ (see Appendix~\ref{Sec:Appendix1}).  If we denote the actual ground state energy of $H_f$ by $E_{\mathrm{gs}}$ then the residual energy converges quadratically~\cite{messiah, suzuki2005residual},
\begin{equation}
    E_{\mathrm{res}} = E-E_{\mathrm{gs}} = \mathcal{O}(1/t_a^2).
\end{equation}
Based on this relation, we propose the following protocol.  We repeat the experiment for an increasing sequence of $t_a$'s and plot the measured $E$'s versus $1/t_a$.  Then we infer the ground state energy $E_{\mathrm{gs}}$ by quadratic regression to the point $1/t_a \rightarrow 0$. The maximum $t_a$ would be determined by the decoherence time on an actual quantum annealer.  

Theoretically, we can do a similar thing for VQE and QITE, $i.e.$, run VQE and QITE repeatedly with different circuit depths (more Trotter steps). However, the residual energy here usually decays exponentially with circuit depth (Trotter steps)~\cite{PhysRevX.10.021067}. Compared with quadratic regression, exponential regression is less stable and robust. In this paper, we therefore concentrate on what can be accomplished
through the polynomial approach, and leave exponential regression for future investigation.

\subsection{Variance Method}\label{sec:variance method}
The variance method is more general in that it can handle evolutions that are characterized by a whole set of parameters (like $\boldsymbol{\theta}$), not just a single parameter (like $t_a$).  We measure and record the variance 
\begin{equation}
    \Delta_{\mathrm{var}} = \langle \psi_f|H_f^2|\psi_f \rangle - |\langle \psi_f|H_f|\psi_f \rangle|^2
\end{equation} 
and the energy $E$ for short annealing times in QA or low-depth circuits in VQE, plot $E(\Delta_{\mathrm{var}})$, then linearly extrapolate the variance to $0$.

To see why this works, note that the final state $|\psi_f\rangle$ can be decomposed as
\begin{equation}\label{psif_dec}
    | \psi_f \rangle = \sqrt{F} | \phi_0 \rangle
    + e^{i\varphi}\sqrt{1-F} | \psi_{\mathrm{ex}} \rangle,
\end{equation}
where $|\phi_0\rangle$ is the exact ground state and $|\psi_{\mathrm{ex}}\rangle$ is a superposition of the exact excited states $\left|\phi_{n}\right\rangle$, \textit{$i.e.$},
\begin{equation}\label{assum}
    \left|\psi_{\mathrm{ex}}\right\rangle=\sum_{j = 1}^{N-1} c_{j}\left|\phi_{j}\right\rangle.
\end{equation}
with $\sum_{j=1}^{N-1} |c_j|^2 = 1 $. $N$ is the dimension of the Hilbert space. $F$ is the fidelity between $| \psi_f \rangle$ and $| \phi_0 \rangle$. When $F\geq1/2$, $E_{\mathrm{res}}$ is upper bounded by $\sqrt{\Delta_{\mathrm{var}}}$~\cite{imoto2021improving}. The variance method is based on the assumption that with the increase of quantum resources, the coefficient $F$ becomes larger, but $|\psi_{\mathrm{ex}}\rangle$ remains essentially unchanged. The condition is expected to hold in the QA when the energy is close to the ground state energy, since the ``leakage" to individual excited states that have avoided level crossings should also be proportional to $1/t_a^2$. In VQE/QITE, it is more difficult to justify this assumption since the parameter set is model-dependent. However, it can be checked simply by confirming the goodness of the linear fit. Under this condition, Refs.~\cite{mizusaki2003precise, shimizu2012variational} show that
\begin{equation}
    \lim_{E_{\mathrm{res}}\rightarrow 0} \Delta_{\mathrm{var}} = \frac{D_2}{D_1}E_{\mathrm{res}},
\end{equation}
where $D_1$ and $D_2$ are constants defined by 
\begin{equation}
    D_{1}=\sum_{j = 1}^{N-1} |c_{j}|^{2}\left(E_{j}-E_{\mathrm{gs}}\right), D_{2}=\sum_{j = 1}^{N-1} |c_{j}|^{2}\left(E_{j}-E_{\mathrm{gs}}\right)^2.
\end{equation}
$E_j$ is the $j$-th excited state eigenenergy of $H_f$. 

Suppose we have several energy-variance points near $(E, \Delta_{\mathrm{var}})$ and use them to do linear extrapolation. The estimated ground state energy is
\begin{equation}\label{eq8}
    E_{\mathrm{extrp}} = E_{\mathrm{gs}} + \frac{E^2_{\mathrm{res}}}{D_2/D_1 - 2E_{\mathrm{res}}},
\end{equation}
so compared with the original data, we reduce the estimation error by
\begin{equation}
    E_{\mathrm{res}}\big(1-E_{\mathrm{res}}/\left(D_2/D_1 - 2E_{\mathrm{res}}\right)\big),
\end{equation}
which is very close to $E_{\mathrm{res}}$ if $E_{\mathrm{res}} \ll D_2/D_1$. Since the second term of Eq.~\ref{eq8} is positive, $E_{\mathrm{extrp}}$ is larger than $E_{\mathrm{gs}}$.

If the quantum evolution is noisy, we approximately describe the final state as
\begin{equation}
    \rho_f = (1-\epsilon')|\psi_f\rangle\langle\psi_f| + \epsilon'\frac{I}{N},
\end{equation}
where $\epsilon'$ is the global depolarizing noise rate. Then, the measured variance is
\begin{equation}
    \Delta_{\mathrm{var}}'=\operatorname{Tr}(\rho_f H_f^2)-\operatorname{Tr}^2(\rho_f H_f),
\end{equation}
the measured energy is $E' = \operatorname{Tr}(\rho_f H_f)$, and the residual energy is $E'_{\mathrm{res}} = E'-E_{\mathrm{gs}}$. 

We assume the hardware noise as a global depolarizing noise channel for several reasons:
\begin{enumerate}
    \item It is believed to be a good approximation to the actual device noise with the circuit is relatively deep~\cite{dalzell2021random}.
    \item Coherent errors ($e.g.$, calibration errors, crosstalk) can be converted to stochastic (depolarizing) noise through the randomized compiling protocol~\cite{hashim2020randomized}.
    \item The global depolarizing channel is relatively easy to analyze.
\end{enumerate}

Suppose $H_f$ is traceless, we have
\begin{equation}
    \lim_{E_{\mathrm{res}}'\rightarrow 0} \Delta_{\mathrm{var}}' = \frac{D_2}{D_1}E_{\mathrm{res}}' + \mathcal{O}(\epsilon'),
\end{equation}
so if $E'_{\mathrm{res}} \ll D_2/D_1$ and $\epsilon'$ is sufficiently small, the linear extrapolation still works. Denote
\begin{equation}
    C' = (D_2/D_1)E_{\mathrm{gs}} + E_{\mathrm{gs}}^2 + \operatorname{Tr}(H^2)/N.
\end{equation}
Now the estimated ground state energy via extrapolation in the presence of noise is
\begin{equation}
\begin{aligned}
    E'_{\mathrm{extrp}} &= E' - \Delta_{\mathrm{var}}' \frac{\partial E'}{\partial \Delta_{\mathrm{var}}'}\\&=E_{\mathrm{gs}} + \frac{E'^2_{\mathrm{res}} - \epsilon'C'}{D_2/D_1 - 2E_{\mathrm{res}}'}.
\end{aligned}
\end{equation}
The interesting thing is that when $\epsilon'$ is an appropriate non-zero small value, $E'_{\mathrm{extrp}}$ is a more precise estimate of $E_{\mathrm{gs}}$ than the noiseless $E_{\mathrm{extrp}}$. The extrapolated energy is even unbiased when $E'^2_{\mathrm{res}} = \epsilon'C'$ strictly holds. Further details of the derivation along with an analysis of the effects of noise can be found in Appendix~\ref{Sec:Appendix2}. 

Suppose the system has $n$ qubits and the Hamiltonian consists of $m$ Pauli observables whose weights are no greater than $k$, 
\begin{equation}
    H_f = \sum_{j=1}^{m} h[j].
\end{equation}
Estimating $\Delta_{\mathrm{var}}$ introduces additional measurement overhead since we need to measure the expectation values of $\mathcal{O}(m^2)$ $2k$-local terms $\{h[i]h[j]\}_{i,j=1,2,\dots,m}$ to estimate
\begin{equation}
    \operatorname{Tr}(\rho_f H_f^2) = \sum_{i=1}^m\sum_{j=1}^m \operatorname{Tr}(\rho_f h[i]h[j]).
\end{equation}

A technique called \textit{classical shadow}, which has been used for hardware error mitigation~\cite{seif2022shadow}, can reduce this cost efficiently~\cite{huang2020predicting, huang2021efficient}. In our protocol, if we construct shadows via random Pauli measurements and use them to predict the local expectation values simultaneously, the number of required measurements are of order $\mathcal{O}(3^k\log (m)/\epsilon^2)$ for estimating local energy terms $\operatorname{Tr}(\rho_f h[j])$ up to additive error $\epsilon$, and are of order $\mathcal{O}(3^{2k}\log (m^2)/\epsilon^2)$ for estimating local variance terms $\operatorname{Tr}(\rho_f h[i]h[j])$ up to $\epsilon$. When $k$ is finite and fixed, the additional measurement overhead is a constant that does not increase with $n$.

Note that neither the infinite-time extrapolation nor the zero-variance extrapolation depends on the QA schedule $f(t)$ and $g(t)$. Our protocols can be applied to QA on arbitrary paths, $e.g.$, QA with a midanneal pausing~\cite{chen2020and}.

\section{Extrapolated Quantum Annealing}\label{Sec:EE_QA}

\begin{figure*}[t]
	\centerline{\includegraphics[width=0.95\textwidth]{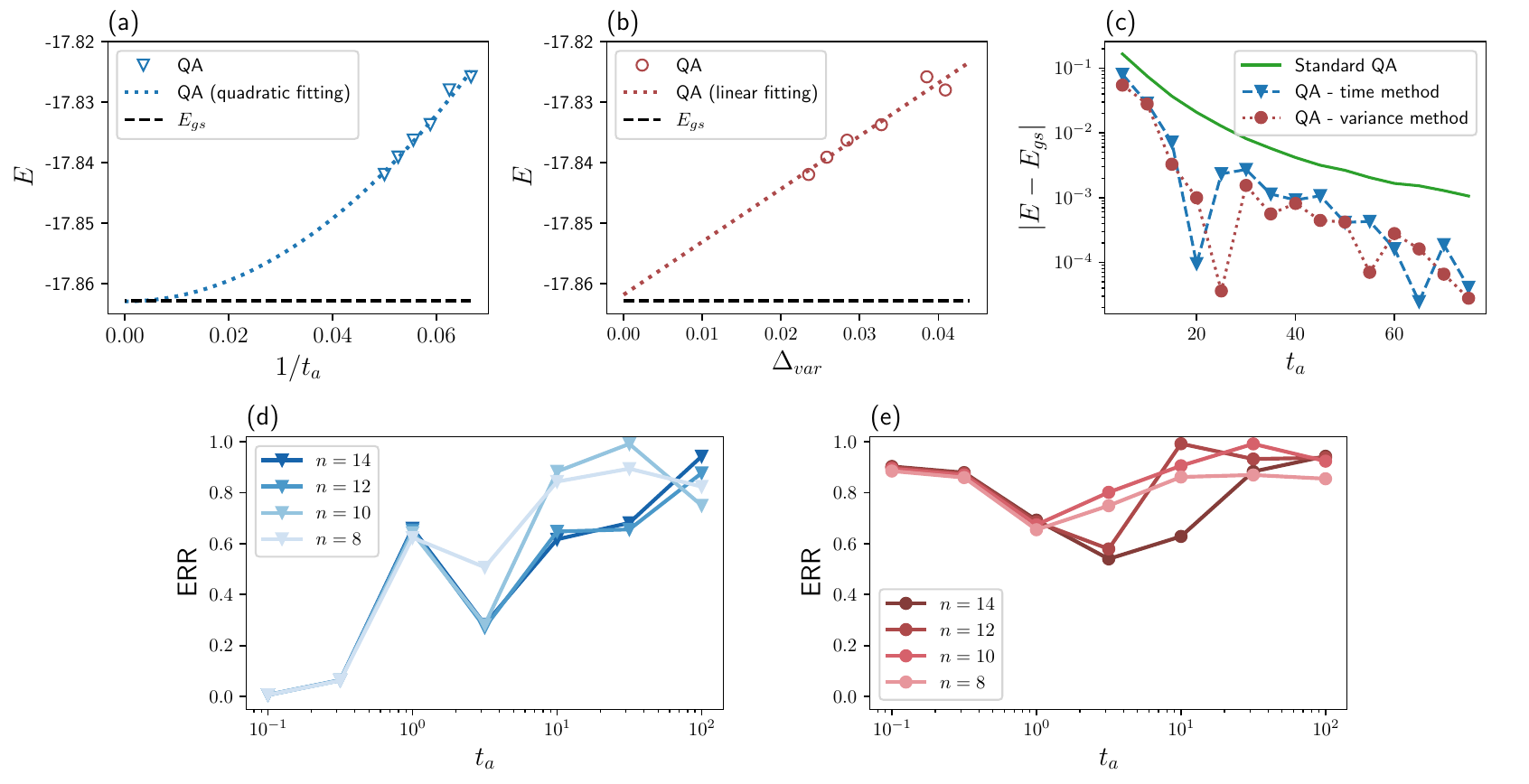}}
	\caption{Extrapolation results obtained in the classical simulations of quantum annealing (QA), showing how very accurate estimates of the ground state energy can be obtained in finite time, even for systems of increasing size. (a) Extrapolating $t_a$ to infinity via quadratic fitting with $t_a = 15,16,17,18,19,20$, $n=14$. The dashed line shows the exact ground state energy. (b) Extrapolating $\Delta_{\mathrm{var}}$ to 0 via linear fitting with $t_a= 15,16,17,18,19,20$, $n=14$. (c) Energy estimation error versus annealing time $t_a$ with $n=14$. (d) Error reduction ratio (ERR), defined in Eq.~\eqref{err_def}, versus system size $n$ with time extrapolation, showing that this method works for relatively large $t_a$. (e) ERR versus system size $n$ with variance extrapolation, showing that this method works for all $t_a$.}
	\label{fig:qa}
\end{figure*}

The workflow of our extrapolation scheme for quantum annealing is illustrated in Algorithm~\ref{alg:QAextp}. Numerical stability could be ensured by using only 6 points for the regression. 

We apply our method to predict the ground state energy of the $n=14$ one-dimensional transverse-field Ising (TFI) model with a periodic boundary condition. The initial Hamiltonian is governed by the transverse field only,
\begin{equation}
    H_{\mathrm{init}} = -\sum_j \sigma_j^x.
\end{equation} 

The final Hamiltonian $H_f$ is the TFI Hamiltonian,
\begin{equation}
    H_{\mathrm{TFI}} = -J \sum_j \sigma_j^z\sigma_{j+1}^z -h \sum_j \sigma_j^x,
\end{equation}
with $J=h=1$. $H_{\mathrm{TFI}}$ is traceless. The system is gapless in the thermodynamic limit. We choose the linear evolution schedule
\begin{equation}
    f(t)=1-t/t_a, \quad g(t)=t/t_a.
\end{equation}

\begin{algorithm}[h]
      \caption{Adaptive energy extrapolation in quantum annealing.}
        \SetKwInOut{Return}{Return}
		\KwIn{Target Hamiltonian $H_f$, update interval $\eta$, convergence tolerance $\epsilon_t$.}
        \KwOut{An accurate estimate of the ground state energy of $H_f$.}
	      $T \leftarrow 1$. \\
	      $E_{\mathrm{extrp}} \leftarrow 0$.\\
	      \While{$E_{\mathrm{extrp}}$ has not converged with tolerance $\epsilon_t$}{Implement QA with 6 evenly spaced annealing times over the interval $[3T/4,T]$. \\
	      Measure the final state and obtain the energies.\\
	      Implement the polynomial regression\\ $\quad\quad E = \alpha_0 + \alpha_1 t_a^{-2}$.\\
	      (Or measure the variances $\Delta_{\mathrm{var}}$ and implement the linear regression $E = \alpha_0 + \alpha_1  \Delta_{\mathrm{var}}$)\\
	      $E_{\mathrm{extrp}} \leftarrow \alpha_0$.\\
	      $T \leftarrow T + \eta$.
	      
	      }
	      \Return{$E_{\mathrm{extrp}}$.}
		\label{alg:QAextp}
\end{algorithm}

The estimation errors with or without extrapolation for different $t_a$ are shown in Fig.~\ref{fig:qa}(c), two detailed extrapolation curves based on time and variance are shown in Fig.~\ref{fig:qa}(a)(b). Both extrapolation schemes can reduce the error by approximately one order of magnitude, $i.e.$, we need much shorter annealing times to estimate the ground state energy to a fixed precision. Considerable speedups can be achieved.

We define the \textit{error reduction ratio} (ERR) as 
\begin{equation}\label{err_def}
    \text{ERR}  = 1 - \frac{|E_{\mathrm{extrp}}-E_{\mathrm{gs}}|}{|E_{\mathrm{std}}-E_{\mathrm{gs}}|},
\end{equation}
where $E_{\mathrm{gs}}$ is the ground state energy of the target Hamiltonian, $E_{\mathrm{std}}$ is the lowest energy obtained by standard QA, $E_{\mathrm{extrp}}$ is the estimated ground state energy obtained via extrapolation. As a goodness criterion, this ratio quantifies to what extent we can mitigate the estimation error. In the ideal case, $\text{ERR}  = 1$, $i.e.$, we mitigate all the estimation errors and obtain the exact ground state energy. ERR as a function of different annealing times and system sizes is shown in Fig.~\ref{fig:qa}(d,e). The scaling with system size is favorable for both schemes. The variance-based scheme can mitigate most errors for all annealing times, and the performance is more stable. The time-based scheme works well for annealing times larger than a critical time.

\begin{figure}[t]
	\centering
	\includegraphics[width=8.5cm]{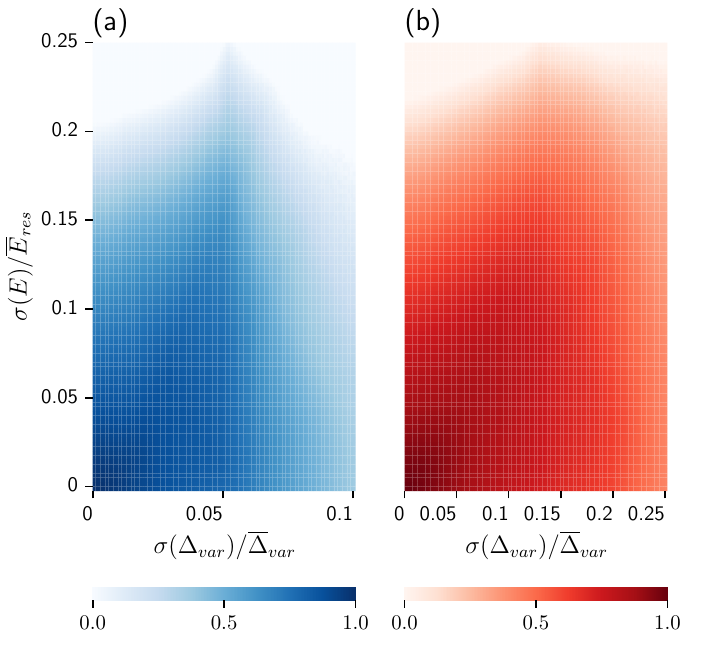}
	\caption{Average Error Reduction Ratio (ERR), Eq.~\eqref{err_def}, in case of measurement and time control errors as a function of standard deviations $\sigma(E)$, $\sigma(\Delta_{\mathrm{var}})$, and $\sigma(t_a)$ for (a) the time method and (b) the variance method. $\overline{t}_a$, $\overline{E}_{\mathrm{res}}$, and $\overline{\Delta}_{\mathrm{var}}$ are the averaged quantities of the fitting data. For both methods there is a broad domain with $\text{ERR} \approx 1$, showing the reasonable robustness of the extrapolation schemes.}
	\label{fig:heatmap}
\end{figure}

Now we consider a more realistic case where the estimates of energy/variance are not exact. This is inevitable on real devices due to readout errors and the limited number of measurement samples. The time control is also not precise. For the 14-qubit TFI model, we run QA with $t_a = 15,16,17,18,19,20$. Suppose $E$, $\Delta_{\mathrm{var}}$, $t_a$ are accompanied by uncorrelated, unbiased, normally distributed precision errors with standard deviations $\sigma(E)$, $\sigma(\Delta_{\mathrm{var}})$, and $\sigma(t_a)$, respectively. We plot the average ERR for different $\sigma(E)$'s, $\sigma(\Delta_{\mathrm{var}})$'s and $\sigma(t_a)$'s in Fig.~\ref{fig:heatmap}. Each pixel is averaged over 10 samples. (If $|E_{\mathrm{extrp}}-E_{\mathrm{gs}}|>|E_{\mathrm{std}}-E_{\mathrm{gs}}|$, we manually set $\text{ERR}  = 0$.) When the standard deviation is smaller than the corresponding quantity by one order of magnitude, the extrapolation method still works and ERR is close to 1. This indicates that the extrapolation method is somewhat stable under measurement and time control errors as long as the error rates are within reasonable bounds. However, we must admit the fact that biased noise is a limitation of our protocol.  In quantum annealing, decoherence would result eventually in thermalization, which always damps the measured energy to 0 (for traceless $H_f$) and increases the measured variance. If the system highly thermalized, the final state is too corrupted to be useful for extrapolation. We leave the problem of dealing with biased noise for future investigation.

\begin{figure*}[t]
	\centerline{\includegraphics[width=0.95\textwidth]{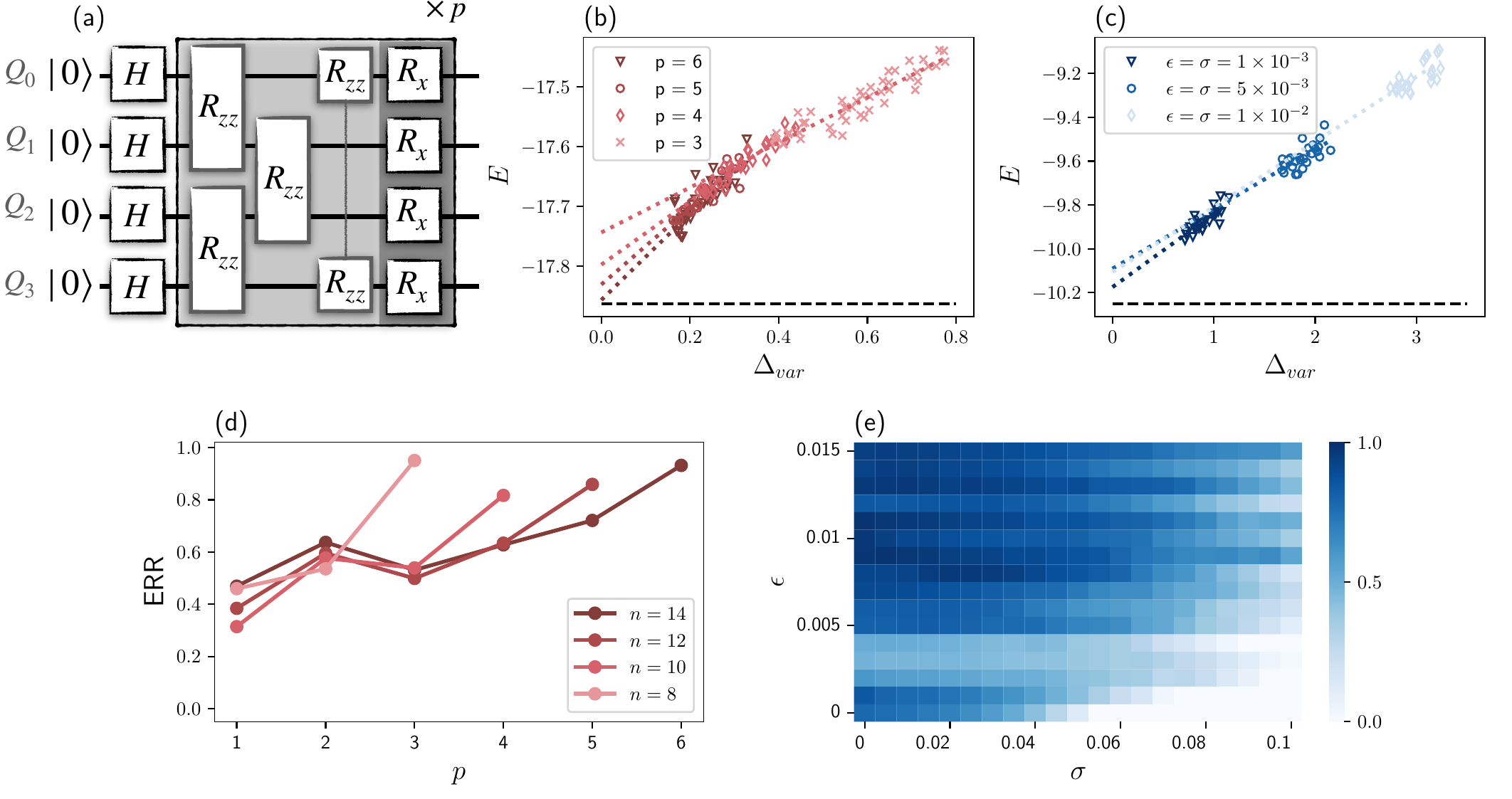}}
	\caption{Circuit structure and extrapolation results obtained in the classical simulations of the variational quantum eigensolver (VQE), showing how very accurate estimates of the ground state energy can be obtained with finite circuit depths. (a) The circuit structure of the Hamiltonian variational ansatz for the 1D transverse-field Ising (TFI) model. (b) Variance extrapolation for different circuit depths with $n = 14$, $\epsilon =0$, $\sigma = 0$. (c) Variance extrapolation for different error rates with $n = 8$. (d) Error Reduction Ratio (ERR) versus depth $p$ for different system sizes, $\epsilon = 0$, $\sigma = 0$. (e) Average ERR as a function of the 2-qubit gate error rate $\epsilon$ and measurement deviation $\sigma$.}
	\label{fig:vqe}
\end{figure*}

\begin{figure*}[t]
	\centerline{\includegraphics[width=0.95\textwidth]{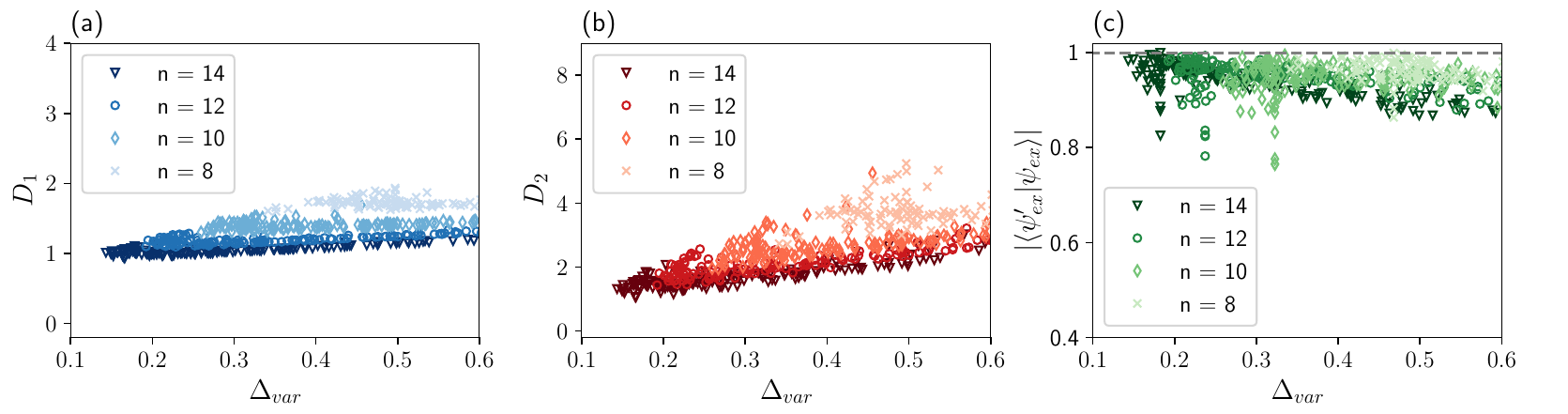}}
	\caption{Numerical verification of the assumption that the excited state superposition $|\psi_{\mathrm{ex}}\rangle$ remains essentially unchanged over the VQE optimization routine. (a) $D_1$ during the optimization. (b) $D_2$ during the optimization. (c) The overlap between $|\psi'_{\mathrm{ex}}\rangle$, the final excited state superposition, and $|\psi^f_{\mathrm{ex}}\rangle$, the excited state superposition during the VQE optimization.}
	\label{fig:overlap}
\end{figure*}


\section{Extrapolated Variational Quantum Eigensolver}\label{Sec:EE_VQE}

VQE is a leading-edge algorithm for ground state preparation using near-term quantum computers. We run parameterized quantum circuits on a quantum computer and use a classical computer to optimize the parameters. There exist different choices of circuits. A well-studied one is the Hamiltonian variational ansatz~\cite{wecker2015progress, wiersema2020exploring}, where the parameterized gates are determined by the target Hamiltonian. Taking the 1D TFI model as an example, we start from the initial state
\begin{equation}
    |\psi_{\mathrm{init}}\rangle = |+\rangle^{\otimes n} = \frac{(|0\rangle + |1\rangle)^{\otimes n}}{\sqrt{2^n}}
\end{equation}
which is the ground state of $-\sum_j \sigma_j^x$, and implement the unitary evolution
\begin{equation}\label{variational circuit}
    U(\boldsymbol{\theta}) = \prod_{l=1}^{p}\exp{[-i\theta_{2l}\sum_j \sigma_j^x]}\exp{[-i\theta_{2l-1}\sum_j \sigma_j^z\sigma_{j+1}^z]}.
\end{equation}
We then iteratively update the $2p$ parameters $\boldsymbol{\theta}$ to minimize the expectation value of the energy of the final state. Previous research indicates that Hamiltonian variational ansatz with depth $p = n/2$ suffices to prepare the ground state of the 1D TFI model~\cite{wierichs2020avoiding, ho2019efficient}. The quantum circuit is illustrated in Fig.~\ref{fig:vqe}(a).

Before enhancing VQE with variance extrapolation, we first numerically verify the assumption that the excited state superposition $|\psi_{\text{ex}}\rangle$ in Eq.~\ref{psif_dec} remains essentially unchanged during the optimization. We denote the excited-state superposition of the fully-optimized state ($i.e.$, the achievable state with the lowest energy) in VQE as $|\psi^f_{\mathrm{ex}}\rangle$ and the excited superposition during the optimization as $|\psi_{\mathrm{ex}}\rangle$. In the following, we set system size $n = 8,10,12,14$ and circuit depth $p=n/2-1$, then sample $10$ different initial points for each case. $D_1$, $D_2$, and the overlaps between $|\psi^f_{\mathrm{ex}}\rangle$ and $|\psi_{\mathrm{ex}}\rangle$ during the optimization are shown in Fig.~\ref{fig:overlap}. $D_1$ and $D_2$ are sightly decreased,  but the changes are within an acceptable range if we abandon points with a large variance. For different system sizes, the overlaps are usually higher than $0.9$ and it gets very close to 1 with the decrease of the variance.

\begin{algorithm}[t]
      \caption{Variance extrapolation in VQE.}
      \SetKwInOut{Return}{Return}
        \KwIn{Target Hamiltonian $H_f$, (noisy) variational quantum circuit of channel $\mathcal{U}_{\boldsymbol{\theta}}$ with random initial parameters $\boldsymbol{\theta}$, initial state $|\psi_{\text{init}}\rangle$, learning rate $\eta$, convergence tolerance $\epsilon_t$, extrapolation ranges $R_{\text{var}}$ and $R_{\text{E}}$, a hardware error mitigation approach $\Lambda$ (optional).}
        \KwOut{An accurate estimate of the ground state energy of $H_f$.}
	      Create empty sets $\mathcal{S}_{\text{E}}$, $\mathcal{S}_{\text{var}}$, $\mathcal{S}^{\Lambda}_{\text{E}}$, and $\mathcal{S}^{\Lambda}_{\text{var}}$.\\
	      $\rho(\boldsymbol{\theta}) \leftarrow \mathcal{U}_{\boldsymbol{\theta}}(|\psi_{\text{init}}\rangle \langle \psi_{\text{init}}|)$.\\
	      $E(\boldsymbol{\theta}) \leftarrow \operatorname{Tr}(\rho(\boldsymbol{\theta})H)$.\\
	      \While{$E(\boldsymbol{\theta})$ has not converged with tolerance $\epsilon_t$}{
	      Estimate the gradient $\nabla E(\boldsymbol{\theta})$ via measurements.\\
	      $\boldsymbol{\theta} \leftarrow \boldsymbol{\theta} - \eta \nabla E(\boldsymbol{\theta})$.\\
	      $\rho(\boldsymbol{\theta}) \leftarrow \mathcal{U}_{\boldsymbol{\theta}}(|\psi_{\text{init}}\rangle \langle \psi_{\text{init}}|)$.\\
	      $E(\boldsymbol{\theta}) \leftarrow \operatorname{Tr}(\rho(\boldsymbol{\theta})H_f)$.\\
	      \If{$E(\boldsymbol{\theta})$ is decreased by 0.01}{
	      Measure $\Delta_{\mathrm{var}}(\boldsymbol{\theta}) = \operatorname{Tr}(\rho(\boldsymbol{\theta})H_f^2) - \operatorname{Tr}^2(\rho(\boldsymbol{\theta})H_f)$.\\
	      Append $E(\boldsymbol{\theta})$ to $\mathcal{S}_{\text{E}}$, $\Delta_{\mathrm{var}}(\boldsymbol{\theta})$ to $\mathcal{S}_{\text{var}}$.\\
	      Apply $\Lambda$ to mitigate the effect of hardware noise and re-estimate the energy $E^{\Lambda}(\boldsymbol{\theta})$ and the variance $\Delta^{\Lambda}_{\mathrm{var}}(\boldsymbol{\theta})$.\\
	      Append $E^{\Lambda}(\boldsymbol{\theta})$ to $\mathcal{S}^{\Lambda}_{\text{E}}$, $\Delta^{\Lambda}_{\mathrm{var}}(\boldsymbol{\theta})$ to $\mathcal{S}^{\Lambda}_{\text{var}}$.\\
	      
	      }}

	      For data pairs $\mathcal{S}_{\text{E}}$-$\mathcal{S}_{\text{var}}$, remove data with energy greater than $\text{min}(\mathcal{S}_{\text{E}}) + R_{\text{E}}$ or with variance greater than $\text{min}(\mathcal{S}_{\text{var}}) + R_{\text{var}}$.\\
	      Implement linear regression on $\mathcal{S}_{\text{E}}$-$\mathcal{S}_{\text{var}}$, obtain $E_{\text{extrp}}$.\\
	      For data pairs $\mathcal{S}^{\Lambda}_{\text{E}}$-$\mathcal{S}^{\Lambda}_{\text{var}}$, remove data with energy greater than $\text{min}(\mathcal{S}^{\Lambda}_{\text{E}}) + R_{\text{E}}$ or with variance greater than $\text{min}(\mathcal{S}^{\Lambda}_{\text{var}}) + R_{\text{var}}$.\\
	      Implement linear regression on $\mathcal{S}^{\Lambda}_{\text{E}}$-$\mathcal{S}^{\Lambda}_{\text{var}}$, obtain $E^{\Lambda}_{\text{extrp}}$.\\
	      \Return{$E_{\text{extrp}}$, $E^{\Lambda}_{\text{extrp}}$.}
	      
		\label{alg:extraVQE}
\end{algorithm}

The variance extrapolation method for VQE works as follows.  During the optimization, both the measured $E$ and the corresponding variance $\Delta_{\mathrm{var}}$ decrease.  We record $E$ and $\Delta_{\mathrm{var}}$ when $E$ is decreased by $0.01$. Denote the energy and variance data sets as $\mathcal{S}_{\text{E}}$ and $\mathcal{S}_{\text{var}}$. After several samples of the initial values and optimization, we find the smallest energy $\text{min}(\mathcal{S}_{\text{E}})$ and the smallest variance $\text{min}(\mathcal{S}_{\text{var}})$, keep only the energy-variance data that satisfy
\begin{equation}
    E < \text{min}(\mathcal{S}_{\text{E}}) + R_{\text{E}}  \quad\text{and}\quad \Delta_{\mathrm{var}}< \text{min}(\mathcal{S}_{\text{var}}) + R_{\text{var}}
\end{equation}
where $R_{\text{var}}$ are $R_{\text{E}}$ are small intervals that in principle should scale with the system size. Here, we choose $R_{\text{var}} = \text{min}(\mathcal{S}_{\text{var}})$, $R_{\text{E}} = 0.5$. After rejecting some data, we implement linear regression to zero for $\Delta_{\mathrm{var}}$ and infer the ground state energy by noting that when the variance vanishes, the energy will be the estimated ground state energy. (In the presence of noise, we modify the extrapolation ranges to $R_{\text{var}} = 1$ and $R_{\text{E}} = 0.5$ ).  The large-variance or large-energy data are abandoned for two reasons: the first one is that the relation between $E$ and $\Delta_{\mathrm{var}}$ is approximately linear only when $\Delta_{\mathrm{var}}$ is close to zero and the ground state is the dominant eigenstate; the second reason is that for some complicated Hamiltonians, the optimization may be stuck to a ``frozen" state whose energy might be low but the variance is quite high, or stuck to an excited state whose variance might be low but the energy is high, and we must avoid the influence of these states. Note that the circuit depth is fixed during the whole process. The workflow of variance extrapolation for VQE is illustrated in Algorithm~\ref{alg:extraVQE}. Our approach is compatible with any hardware error mitigation approach.

We apply the extrapolation-assisted VQE to the TFI Hamiltonian with a periodic boundary condition and consider the noiseless situation first. We sample 10 initial points and update the parameters with the BFGS \cite{fletcher} algorithm. The extrapolated results for $n = 14$, $p=3,4,5,6$ are shown in Fig.~\ref{fig:vqe}(b). Our method can reduce the estimation error by 55\%, 64\%, 76\%, and 94\%, respectively. Depth-3 VQE with extrapolation outperforms depth-5 VQE, and depth-4 VQE with extrapolation outperforms depth-6 VQE. Since training a shallower variational quantum circuit requires fewer training epochs, extrapolation-assisted VQE also saves time. ERR as a function of $n$ and $p$ is illustrated in Fig.~\ref{fig:vqe}(d). Unsurprisingly, ERR increases with circuit depth $p$ since the relation between $E_{\mathrm{res}}$ and $\Delta_{\mathrm{var}}$ is only approximately linear for small $\Delta_{\mathrm{var}}$'s. Fig.~\ref{fig:vqe}(d) suggests that our method still works for larger system sizes. It is worth mentioning that in the extrapolation-assisted VQE, the local minima problem is less severe since even if the optimization is trapped to a local minimum eventually, the energies and variances recorded suffice to predict the ground state energy accurately.

Now we show that our method is still efficient in the advent of noise. Suppose every $2$-qubit $Rzz$ gate $\exp{[-i\theta \sigma_j^z\sigma_{j+1}^z]}$ in Eq.~\eqref{variational circuit} is accompanied by a local depolarizing channel with noise rate $\epsilon$. For the output state, we estimate $E$ and $\Delta_{\mathrm{var}}$ with normally distributed errors (we assume the estimating standard deviations of $E$ and $\Delta_{\mathrm{var}}$ are both $\sigma$). For $n = 8$, $p = 3$, we implement variance extrapolation for the noisy VQE, and the results are shown in Fig.~\ref{fig:vqe}(c). As expected, the minimum energy grows rapidly with increasing error rate. The estimation error of $E_{\mathrm{gs}}$ here is caused by both algorithmic and hardware deficiencies. Encouragingly, variance extrapolation can always reduce the majority of the estimation error and give an accurate estimate of $E_{\mathrm{gs}}$. For $\epsilon = \sigma = 0.001, 0.005,$ and $0.01$, our method can reduce the estimation error by 73\%, 73\%, and 85\%, respectively. In the last case, the ground state energy, the lowest energy achieved by VQE, and the extrapolated energy are
\begin{equation}
    E_{\mathrm{gs}} = -10.25, \mkern7mu\text{min}(\mathcal{S}_{\text{E}}) = -9.30,\mkern7muE_{\mathrm{extrp}} = -10.19.
\end{equation}
This indicates that one can directly use the extrapolation technique for NISQ devices since the two-qubit gate noise rate on a state-of-the-art quantum computer is approximately 0.01~\cite{ai2021exponential}. Within the range $(0,0.01)$, the average ERR even increases with the noise rate. This phenomenon agrees with our analysis in Sec.~\ref{sec:variance method} that a small noise can play a positive role. Further, we plot the average ERR for different $\epsilon$'s and $\sigma$'s in Fig.~\ref{fig:vqe}(e). Our method is reasonably robust against noise, and it can serve as a tool for algorithmic and hardware error mitigation.

\begin{figure*}[t]
	\centerline{\includegraphics[width=0.95\textwidth]{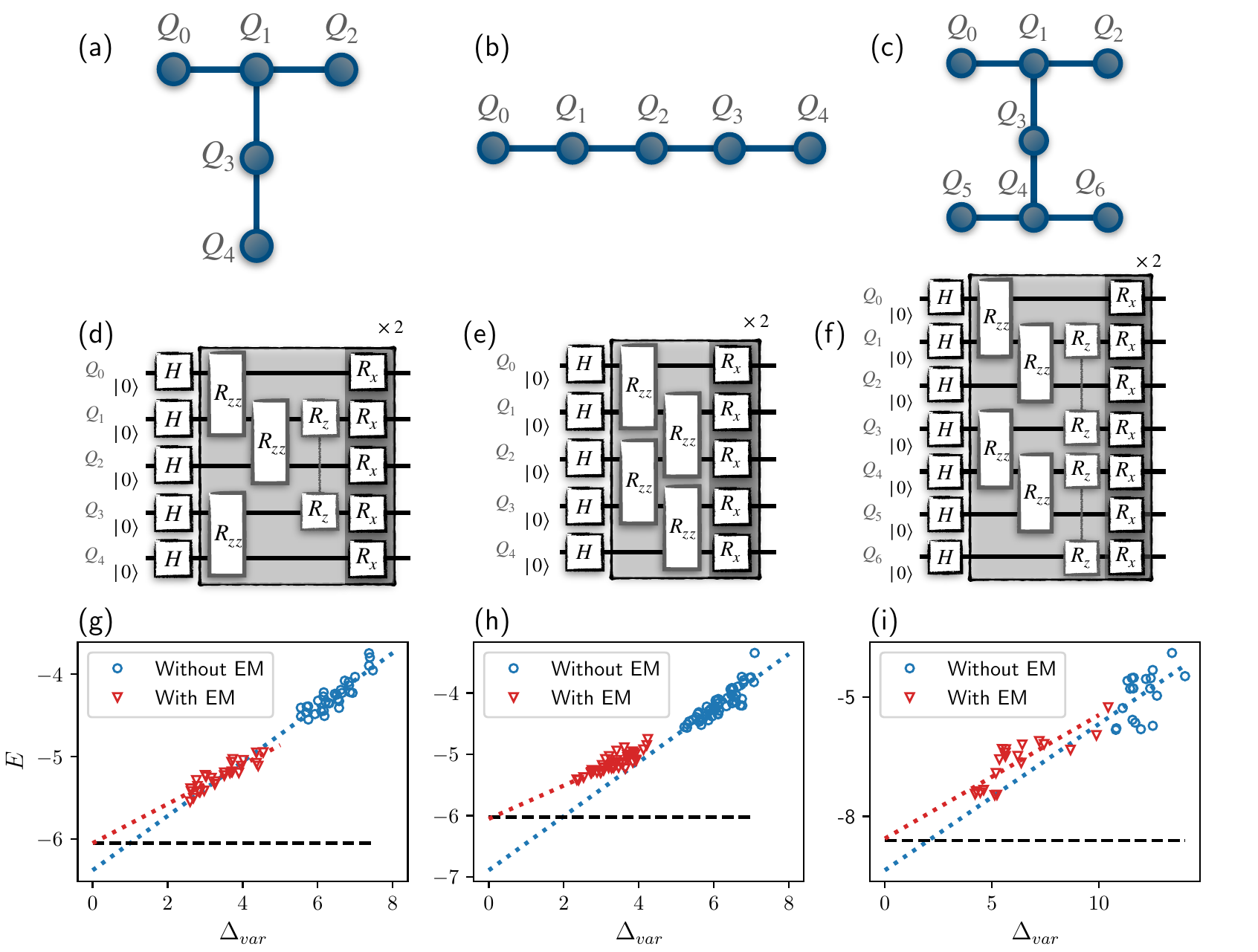}}
	\caption{Experimental results of the variational quantum eigensolver on real IBM quantum computers, showing how an accurate estimate of the ground state energy can be obtained via extrapolation. (a-c) The hardware connectivities of $ibmq\_quito$, $ibmq\_manila$ and $ibmq\_perth$. Filled circles represent the qubits. (d-f) The hardware-efficient variational quantum circuits to prepare the ground state of the transverse-field Ising model for $ibmq\_quito$, $ibmq\_manila$ and $ibmq\_perth$. (g-i) Linear extrapolation with measured energies and variances on $ibmq\_quito$, $ibmq\_manila$, and $ibmq\_perth$, respectively. The blue circles represent experimental data without error mitigation (EM), the red triangles represent data with errors mitigated by ``virtual distillation". The dashed line shows the exact ground state energy $E_{\mathrm{gs}}$.}
	\label{fig:ibmq}
\end{figure*}

It is worth noting that the extrapolation method works for VQE with different ansatzes. To better demonstrate the practicability of our approach, we implement VQE experiments with a hardware-efficient circuit on real IBM quantum devices. The machines we use are $ibmq\_quito$ (5 qubits), $ibmq\_manila$ (5 qubits), and $ibmq\_perth$ (7 qubits)~\cite{ibmq_quito, ibmq_manila, ibmq_perth}. Their average CNOT error rates are $1.065 \times 10^{-2}$, $9.885 \times 10^{-3}$, and $1.158 \times 10^{-2}$, respectively. The connectivities are given in Fig.~\ref{fig:ibmq}(a-c). Still, we consider the transverse-field Ising model
\begin{equation}
    H_{\mathrm{TFI}} = - J\sum_{<ij>} \sigma_i^z\sigma_j^z - h\sum_j\sigma_j^x
\end{equation}
with $\langle ij \rangle$ denoting adjacent qubits, $J=h=1$, and use Hamiltonian variational quantum circuits illustrated in Fig.~\ref{fig:ibmq}(d-f) to prepare the ground states. These circuits with $p=2$ are not capable of preparing the exact ground states even without gate noise.

We start from $|0\rangle^{\otimes n}$ and apply Hadamard gates to prepare $|\psi_{\mathrm{init}}\rangle =|+\rangle^{\otimes n}$. The initial parameters are uniformly randomly sampled from $[0, 2\pi)$ and updated with a classical optimizer. The energies and variances are measured on real quantum machines without readout error mitigation. Here we specifically use a quantum error mitigation technique called \textit{virtual distillation}~\cite{koczor2021exponential, PhysRevX.11.041036}. The main idea of virtual distillation is to perform joint measurements on multiple copies of $\rho$ to estimate the expectation value on
\begin{equation}
    \rho_{\text{EM}} = \frac{ \rho_f^{M}}{\operatorname{Tr}\left(\rho_f^{M}\right)}
\end{equation}
with $M \geq 2$. The larger the value of $M$, the purer the state $\rho_{\text{EM}}$. In the following we fix $M=2$ and set $R_{\text{E}}=R_{\text{var}}=2$ ($ibmq\_quito$), $R_{\text{E}}=R_{\text{var}}=2$ ($ibmq\_manila$), $R_{\text{E}}=R_{\text{var}}=6$ ($ibmq\_perth$). Due to hardware constraints, we cannot couple qubits in different copies so instead we measure an orthonormal Hermitian operator basis with $1024$ shots each,  and compute the maximum likelihood density matrix $\rho$~\cite{Smolin_2012}, then compute $\rho_{\text{EM}}$ and the error-mitigated expectation values $E^{\Lambda}(\boldsymbol{\theta})$, $\Delta^{\Lambda}_{\mathrm{var}}(\boldsymbol{\theta})$ classically. Variance extrapolation is effective for data either with or without error mitigation. The results are shown in Fig.~\ref{fig:ibmq}(g-i).  Clearly, we can mitigate the majority of the estimation errors in all cases. For the original data without error mitigation, the extrapolated energy is slightly lower than the ground state energy, which is consistent with our theory discussed in Sec.~\ref{sec:variance method}. For the data with error mitigation, the extrapolated energy gets very close to the ground state energy, $e.g.$, $E_{\text{gs}}$. The ground state energies, the lowest energies achieved by VQE, and the extrapolated energies with error-mitigated data ($\mathcal{S}^{\Lambda}_{\text{E}}$, $\mathcal{S}^{\Lambda}_{\text{var}}$) in the three experiments are
\begin{equation}
\begin{aligned}
    &ibmq\_quito: \\&E_{\mathrm{gs}} = -6.05, \mkern7mu\text{min}(\mathcal{S}^{\Lambda}_{\text{E}}) = -5.55,\mkern7muE^{\Lambda}_{\mathrm{extrp}} = -6.05;\\
    &ibmq\_manila: \\&E_{\mathrm{gs}} = -6.03, \mkern7mu\text{min}(\mathcal{S}^{\Lambda}_{\text{E}}) = -5.42,\mkern7muE^{\Lambda}_{\mathrm{extrp}} = -6.05;\\
    &ibmq\_perth: \\&E_{\mathrm{gs}} = -8.61, \mkern7mu\text{min}(\mathcal{S}^{\Lambda}_{\text{E}}) = -7.39,\mkern7muE^{\Lambda}_{\mathrm{extrp}} = -8.55.
\end{aligned}
\end{equation}
In addition to virtual distillation, variance extrapolation is theoretically compatible with other quantum error mitigation techniques such as the learning-based approach~\cite{PRXQuantum.2.040330, Sopena_2021}.

\section{Extrapolated Quantum Imaginary Time Evolution}\label{Sec:EE_QITE}

\begin{figure}[h!]
	\centering
	\includegraphics[width=7cm]{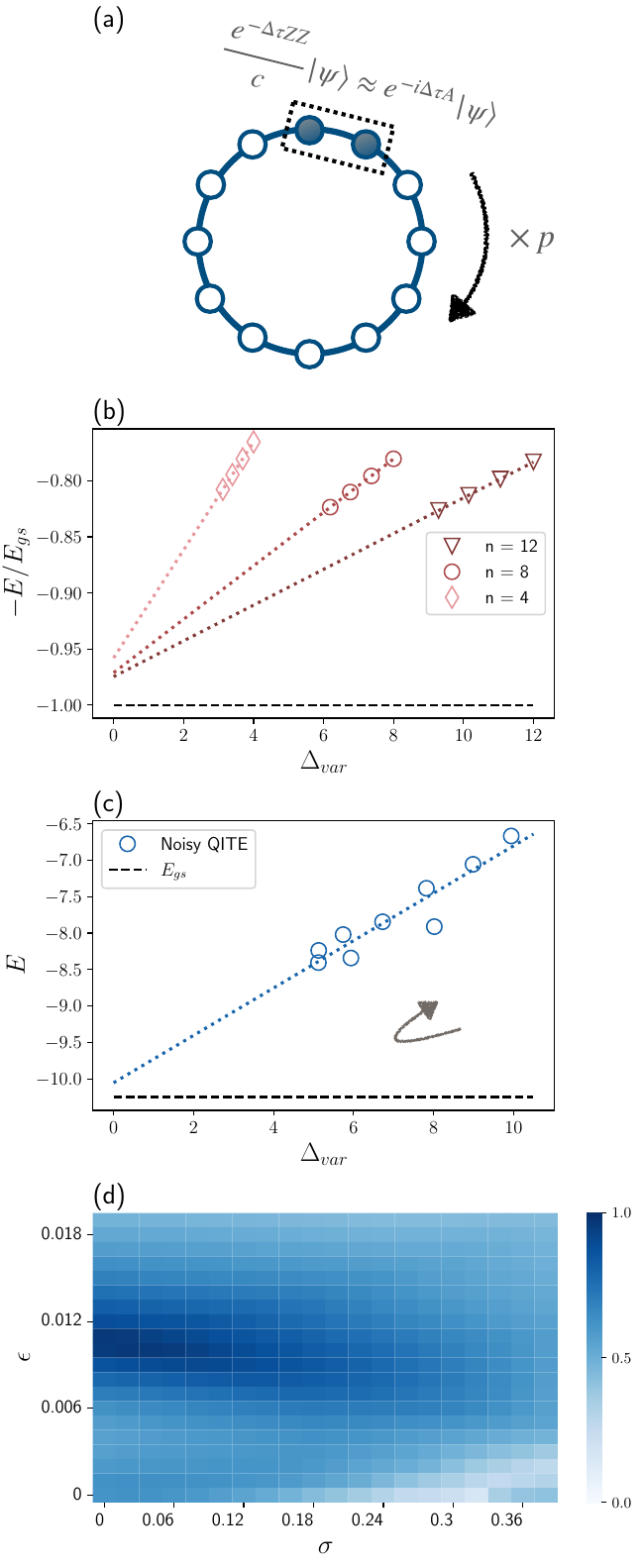}
	\caption{Schematic and extrapolation results of the quantum imaginary time evolution (QITE) algorithm, showing how better estimates of the ground state energy can be obtained with finite Trotter steps. (a) Schematic of QITE for the 1-D transverse-field Ising (TFI) model. (b) Variance extrapolation for different system sizes with 3 Trotter steps, $\epsilon =0$, $\sigma = 0$. (c) Variance extrapolation for the noisy QITE with $n = 8$, $\epsilon = \sigma =0.01$.  (d) Average error reduction ratio as a function of the 2-qubit gate error rate $\epsilon$ and measurement deviation $\sigma$.}
	\label{fig:qite}
\end{figure}

QITE is a NISQ algorithm inspired by the numerical technique of imaginary time evolution~\cite{motta2020determining}. Although QITE was initially proposed for ground state preparation, it has a variety of other potential applications, such as generating thermal states~\cite{zeng2021variational} and simulating open quantum systems~\cite{PRXQuantum.3.010320}. Here we restrict our attention to estimating ground-state energies. 

Consider a $2$-local Hamiltonian $H_f=\sum_{j=1}^m h[j]$ with each local term $h[j]$ acting on two neighboring qubits. For any initial state $|\psi_{\mathrm{init}}\rangle$, as long as the ground subspace of $H_f$ is not orthogonal to $|\psi_{\mathrm{init}}\rangle$, the final state after long-time imaginary time evolution 
\begin{equation}
    |\psi_f\rangle = \lim _{\beta \rightarrow \infty} e^{-\beta H_f}\left|\psi_{\mathrm{init}}\right\rangle
\end{equation}
falls into the ground subspace of $H_f$. This process can be simulated by the $p$-step Trotter decomposition
\begin{equation}\label{trotter}
    e^{-\beta H_f}=\left(e^{-\Delta \tau h[1]} e^{-\Delta \tau h[2]} \ldots e^{-\Delta \tau h[m]}\right)^{p}+\mathcal{O}(\Delta \tau ^2),
\end{equation}
with $\Delta \tau = \beta/p$ being the step interval. Since quantum computers can only implement unitary evolutions, the main idea of the QITE algorithm is to replace each non-unitary evolution $e^{-\Delta \tau h[j]}$ with a $D$-local unitary evolution $e^{-i\Delta \tau A[j]}$ such that the states after these two processes are very close:
\begin{equation}\label{local_approx}
    \frac{e^{-\Delta \tau h[j]}}{\sqrt{\langle\psi|e^{-2 \Delta \tau h[j]}| \psi\rangle}}|\psi\rangle \approx e^{-i \Delta \tau A[j]}|\psi\rangle.
\end{equation}
The optimal $A[j]$ can be determined via measurements on $|\psi\rangle$ and solving a linear equation classically~\cite{motta2020determining}. For larger system sizes, we need larger $\beta$ values. The number of measurements required to perform QITE does not scale favorably with system size.

There are three challenges for the QITE algorithm. The first one is the Trotter error induced by Trotterization. On NISQ devices, we need to choose a small $p$ to obtain reliable results. For these small value of $p$, the step interval $\Delta \tau$ is large, and the corresponding Trotter error is non-negligible.  The second one is the local approximation error: we have to use a small-size local operator $e^{-i \Delta \tau A[j]}$ to approximate each non-unitary $e^{-\Delta \tau \hat{h}[j]}$ (we set $D=2$ in the following numerical tests), so relation~(\ref{local_approx}) is not accurate. The third one is the hardware noise. Errors occurring at an early stage of the calculation may prevent us from finding the correct $A[j]$ in a later step. All these errors accumulate with circuit depth. As a result, QITE is highly sensitive to errors.

We again consider the TFI model and use the variance method to enhance QITE. The initial state is $|\psi_{\mathrm{init}}\rangle = |+\rangle^{\otimes n}$. Specifically, we record both the energy and the variance after each Trotter step. After several steps, we implement extrapolation with the recorded data. Fig.~\ref{fig:qite}(b) shows some noiseless extrapolation results with only $3$ Trotter steps. Even very shallow circuits can yield reasonable estimates. The variance method can reduce $78\%$,$84\%$,$85\%$ of the estimation error for $n=4,8,12$, respectively. The mitigating effect is scalable. Now suppose each $2$-qubit operator $e^{-i \Delta \tau A[j]}$ is accompanied by depolarizing noise with the rate $\epsilon$, and the estimating standard deviations of $E$ and $\Delta_{\mathrm{var}}$ are both $\sigma$.  Fig.~\ref{fig:qite}(c) plots a detailed example for $\epsilon =\sigma = 0.01$, where the energy decreases first, then increases because the errors become dominant. Here, our method reduces the estimation error by $89\%$. The actual ground state energy, the lowest energy achieved by QITE, and the extrapolated energy are
\begin{equation}
E_{\mathrm{gs}} = -10.25, \mkern7mu\text{min}(\mathcal{S}_{\text{E}}) = -8.41,\mkern7muE_{\mathrm{extrp}} = -10.05.
\end{equation}
Fig.~\ref{fig:qite}(d) shows the average ERR for different $\epsilon$ and $\sigma$. As we analyzed in Sec.~\ref{sec:variance method}, the extrapolation method performs especially well for a small non-zero $\epsilon$. Encouragingly, that value here is $\sim 0.01$, which matches the noise rates of current NISQ computers.

\section{Conclusions and Outlook}
\label{Sec:Conclusions}



In this article we have shown how the zero-variance and infinite--time 
extrapolation schemes can be used to improve the results of 
quantum optimization calculations on noisy intermediate-scale quantum 
(NISQ) devices.
In the case of infinite-time extrapolation, we have established the 
asymptotic behavior found in quantum annealing (QA), and
used numerical simulations to demonstrate the effectiveness of the method and 
its reasonable robustness against measurement noise.
In the case of zero-variance extrapolation, we have considered 
applications to quantum annealing (QA), the variational quantum eigensolver (VQE) and the quantum imaginary-time evolution (QITE).
We have used numerical simulations to show that variance extrapolation 
leads to improvements in the results found with each of these algorithms, 
even in the presence of finite gate noise, without leading to any 
additional overhead in circuit depth.
We have also explicitly demonstrated the effectiveness of 
zero-variance extrapolation in a VQE calculation carried out 
on three IBM quantum computers.

%
%


It is conceivable that other NISQ algorithms like quantum approximate optimization algorithm (QAOA) can similarly benefit from this method. One may use the extrapolation technique to accurately estimate excited-state energies in excited-state-preparing algorithms such as the subspace-search VQE~\cite{nakanishi2019subspace}. In addition, since the adiabatic model and the circuit model are polynomially equivalent~\cite{aharonov2008adiabatic}, any quantum circuit can be translated to the ground state problem via the Feynman-Kitaev circuit-to-Hamiltonian construction~\cite{kitaev2002classical}, so the time and variance-based extrapolation may benefit other areas of quantum computing as well, $e.g.$, digital quantum simulation. Energy extrapolation is a scalable technique that pushes NISQ algorithms to more practical applications. 

In quantum chemistry, many molecular properties we are concerned about (e.g., the dipole moment, the moment of inertia) can be written as derivatives of the ground state energy~\cite{helgaker2014molecular}. Therefore, one can estimate these quantities more accurately with energy extrapolation. In quantum information and condensed matter physics, we are also interested in knowing ground state properties like phase and correlators. Given various applications of variance extrapolation in quantum Monte Carlo like estimating quadrupole moments~\cite{shimizu2010novel}, there is reason to believe that extrapolation can as well help estimate these physical quantities. How to combine the extrapolation method with other techniques to obtain more pieces of physical information is worth studying further.

\acknowledgments

We thank Rico Pohle, Xiang-Bin Wang, Jinfeng Zeng, and Junan Lin for helpful discussions and suggestions. We acknowledge the use of IBM Quantum services for this work. CC and BZ are supported by General Research Fund (no. GRF/16300220).




\appendix
\section{Quadratic convergence of the residual energy}\label{Sec:Appendix1}
Here we give the justification for the statement that
\begin{equation}
\label{eq:eres}
    E_{\mathrm{res}} = \frac{A}{t_a^2} + \mathcal{O}(t_a^{-3}),
\end{equation}
find an expression for the constant $A$, and give a criterion for when the $1/t_a^3$ term can be neglected.

We have an $N$-dimensional Hilbert space and a time-dependent Hamiltonian with instantaneous eigenstates $| \phi(t) \rangle $:
\begin{equation}
\label{eq:ins}
    H(t) \, | \phi_j(t) \rangle = \varepsilon_j(t) \, | \phi_j(t) \rangle
\end{equation}
where $j=0,1,...N-1$. We assume that there are no symmetry-related level crossings so that the ordering of the states is preserved in the evolution.  A general state $| \psi(t) \rangle$ that satisfies the Schr\"{o}dinger equation may be expanded in the diabatic basis as
\begin{equation}
    | \psi(t) \rangle = 
    \sum_{j=0}^{N-1}
    a_j(t) \exp \left[-i \int_0^t \varepsilon_j(t') dt'\right] \, \,
    | \phi_j(t) \rangle  
\end{equation}

where the coefficients satisfy
\begin{equation}
\begin{aligned}
  \partial_t a_k(t) = - &\sum_{j=0}^{N-1} a_j(t)  
  \exp \left\{-i \int_0^t \left [\varepsilon_j(t') - \varepsilon_k(t') \right] dt' \right\}
  \\&\langle \phi_k(t) | \partial_t H | \phi_j(t) \rangle \,
  [\varepsilon_j(t') - \varepsilon_k(t') ] ^{-1}.
\end{aligned}
\end{equation}

If we start the system in the ground state $| \phi_0(t=0) \rangle$ and neglect multiple scattering we can set $a_j(t) = \delta_{0j}$ on the right-hand side of this equation and integrate over time to obtain
\begin{equation}
\begin{aligned}
\label{eq:ak1}
  a_k(t) = \int_0^t &dt'
  \exp \left\{i \int_0^{t'} \left [\varepsilon_k(t'') - \varepsilon_0(t'') \right] dt'' \right\}
  \\&\langle \phi_k(t') | \partial_{t'} H(t') | \phi_0(t') \rangle \,
  [\varepsilon_k(t') - \varepsilon_0(t') ] ^{-1}.  
\end{aligned}
\end{equation}
By differentiating Eq. \eqref{eq:ins} and substituting the result we obtain the equivalent and slightly simpler form
\begin{equation}
\begin{aligned}
\label{eq:ak2}
  a_k(t) = - \int_0^t &dt'
  \exp \left\{i \int_0^t \left [\varepsilon_k(t') - \varepsilon_0(t') \right] dt' \right\} 
   \\&\langle \phi_k(t') | \partial_{t'} \phi_0(t') \rangle.  
\end{aligned}
\end{equation}
Eqs.~\ref{eq:ak1} and \ref{eq:ak2} are standard results ~\cite{migdal}.

To isolate the dependence on $t_a$ it's now convenient to change variables to $s=t/t_a$.  $\varepsilon_0(s=1) = E_{\mathrm{gs}}$ is the final ground state energy. The final state of the evolution is $|\psi(s=1) \rangle$ while the actual final ground state is $|\phi_0(s=1) \rangle$.  The probability to find the system in the $j$th excited state at the end is $p_j = |a_j(s=1)|^2 $, so the that probability of failure of the algorithm is the sum of the  $|a_j(s=1)|^2 $ from $j=1$ to $j=N-1$.

The $a_j(s=1)$ are given by
\begin{equation}
\begin{aligned}
\label{pj_ana}
  a_j(s=1) = - \int_0^1 &ds'
  \exp \left\{i t_a \int_0^{s'} \left [\varepsilon_j(s'') - \varepsilon_0(s'') \right] ds'' \right\} 
   \\&\langle \phi_j(s') | \partial_{s'} \phi_0(s') \rangle.  
\end{aligned}
\end{equation}
This may be written as 
\begin{equation}
  a_j(s=1) = - \int_0^1 ds'
  \exp \left[ i t_a F_j(s') \right] G_j(s'),
\end{equation}
where 
\begin{equation}
    F_j(s) = \int_0^{s} \left [\varepsilon_j(s') - \varepsilon_0(s') \right] ds'  
\end{equation}
and
\begin{equation}
    G_j(s) =  \langle \phi_j(s) | \partial_{s} \phi_0(s) \rangle = - \langle \phi_j(s) | \partial_s H | \phi_0(s) \rangle \
\end{equation}
and since $F_j$ and $G_j$ are continuous the Riemann-Lebesgue Lemma guarantees that this expression vanishes as $t_a \rightarrow \infty$ for all $j$. 

\begin{figure}[t]
	\centering
	\includegraphics[width=7cm]{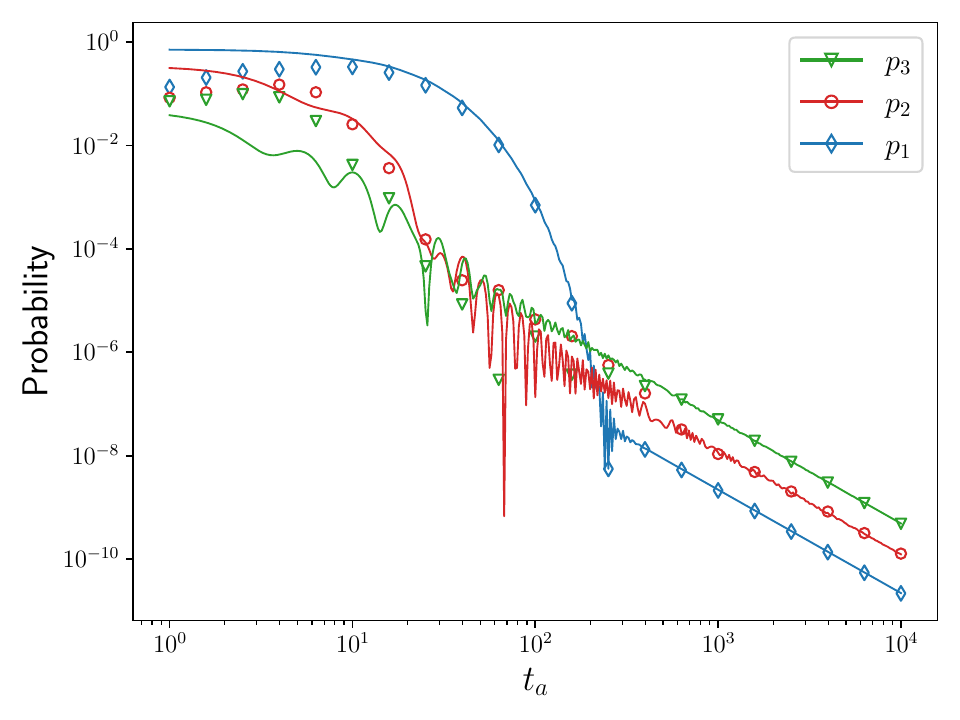}
	\caption{The probability to find the system in the $j$-th excited state at the end of quantum annealing. The markers represent the numerical results, while the lines represent the analytical results of Eq.~ \ref{pj_ana}.}
	\label{fig:pj}
\end{figure}

We take a simple random-field Ising Hamiltonian as an example.  The probabilities for various states are shown in Fig.~\ref{fig:pj}.  Numerical and analytical results match well when $t_a > 10$.

Since $F_j'(s)>0$ for all $s$ we can use integration by parts to develop an asymptotic expansion in $1/t_a$~ \cite{bender} :
\begin{equation}
\begin{aligned}
  a_j(s=1) = &- \frac{1}{it_a} \left[e^{it_a F_j(s)} 
  \frac{d}{ds} \left( \frac{G_j(s)}{F_j'(s)} \right) \right]^1_0 \\&+ \frac{1}{(it_a)^2} \left\{e^{it_a F_j(s)} 
  \frac{d}{ds} \left[ \frac{1}{F_j'(s)} \frac{d}{ds} \left( \frac{G_j(s)}{F_j'(s)} \right) \right] \right\}^1_0 \\&+ \mathcal{O}(\frac{1}{t_a^3})
\end{aligned}
\end{equation}

The residual energy in this approximation is 
\begin{equation}
    E_{\mathrm{res}} = 
    \sum_{j=1}^{N-1}
    |a_j(s=1)|^2
    \left[ \varepsilon_j(s=1)-\varepsilon_0(s=1) \right],
\end{equation}
The leading term of $a_j(s=1)$ is $\mathcal{O}(1/t_a)$, so the leading term of $E_{\mathrm{res}}$ is $\mathcal{O}(1/t^2_a)$. The coefficient $A$ in Eq.~\eqref{eq:eres} is
\begin{equation}
    A =  \sum_{j=1}^{N-1}
    \left| \left[e^{it_a F_j(s)} 
  \frac{d}{ds} \left( \frac{G_j(s)}{F_j'(s)} \right) \right]^1_0 \right|^2 
   \left[ \varepsilon_j(s=1)-\varepsilon_0(s=1) \right].
\end{equation}
In an asymptotic expansion, the first term dominates when the second term is much smaller, which leads to the criterion
\begin{equation}
    t_a \gg
 \frac{\left| \left\{ e^{it_a F_j(s)} 
  \frac{d}{ds} \left[ \frac{1}{F_j'(s)} \frac{d}{ds} \left( \frac{G_j(s)}{F_j'(s)} \right) \right] \right\}^1_0 \right |}{
  \left | \left[e^{it_a F_j(s)} 
  \frac{d}{ds} \left( \frac{G_j(s)}{F_j'(s)} \right) \right]^1_0 \right|}.
\end{equation}
This must hold for all $j$.

Since $F'_j(s) = \varepsilon_j(s)-\varepsilon_0(s) $ we see that the criterion does depend on the behavior of the gap between the ground state energy and the excited state energies, albeit only at the endpoints of the evolution.  The criterion becomes more stringent as the gaps decrease, as is expected.  Since this also applies to $s=0$, this would suggest using an initial Hamiltonian with a large gap. We leave this problem for future investigation.

\section{Linear convergence of the energy variance}\label{Sec:Appendix2}
In this section, we present the criterion for the linearity of the variance and use it to justify the extrapolation procedure. There is a set of exact eigenstates $| \phi_j \rangle$ of $H_f$: 
\begin{equation}
    H_f | \phi_j \rangle = E_j | \phi_j \rangle.
\end{equation}
Our goal is to determine $E_0 = E_{\mathrm{gs}}$ by a sequence of variational calculations.
The variational wavefunction $| \psi_f \rangle $ at any stage of the sequence can be written as 
\begin{equation}
    | \psi_f \rangle = \sqrt{F} | \phi_0 \rangle
    + e^{i\varphi}\sqrt{1-F} | \psi_{\mathrm{ex}} \rangle,
\end{equation}
where $| \phi_0 \rangle$ is the exact ground state and $| \psi_{\mathrm{ex}} \rangle$ is a sum of the exact excited eigenstates $| \phi_j \rangle$ of $H_f$:
\begin{equation}
    | \psi_{\mathrm{ex}} \rangle = \sum_{j=1}^{N-1} c_j  
    | \phi_j \rangle
\end{equation}
with $\sum_{j=1}^{N-1} |c_j|^2 = 1 $.  $F$ is the fidelity.
We follow Ref.~\cite{shimizu2012variational} and define two energy moments
\begin{equation}
    D_1 = \sum_{j=1}^{N-1} |c_j|^2 (E_j-E_{\mathrm{gs}})   
\end{equation}
and 
\begin{equation}
    D_2 = \sum_{j=1}^{N-1} |c_j|^2  (E_j-E_{\mathrm{gs}})^2 .   
\end{equation}
Then some manipulation leads to the equations
\begin{equation}
    E-E_{\mathrm{gs}} = E_{\mathrm{res}} = (1-F) D_1
\end{equation}
and 
\begin{equation}
    \Delta_{\mathrm{var}} (s=1) = (1-F) D_2  - (1-F)^2D_1^2.
\end{equation}
In the variance extrapolation we perform a sequence of calculations in which $F$ increases but is not measured while $E(F)$ and $\Delta_{\mathrm{var}}(F)$ are measured.  Thus we wish to eliminate $F$ from these equations.  When this is done we find
\begin{equation}
\label{eq:var_e}
    \Delta_{\mathrm{var}} = \frac{D_2}{D_1} E_{\mathrm{res}} - E_{\mathrm{res}}^2
    =\frac{D_2}{D_1} (E-E_{\mathrm{gs}}) - (E-E_{\mathrm{gs}})^2.
\end{equation}
Now we assume that $|c_j|^2$, the relative weights of the excited states, and hence also the $D_{1,2}$, do not vary significantly with $F$ near $F=1$.  Then we see that the dependence of $\Delta_{\mathrm{var}}$ on $E_{\mathrm{res}}$ is parabolic and linear when $E-E_{\mathrm{gs}} \approx 0$.  This criterion for the validity of the linear approximation used in the main text is that 
\begin{equation}
\label{eq:eres1}
    E_{\mathrm{res}} = E-E_{\mathrm{gs}} \ll D_2/D_1.
\end{equation} 
One cannot easily calculate $D_{1,2}$, since they depend on the structure of $H_f$ as well as on the details of the variational procedure.  In practice, however, this is not necessary.  One plots $\Delta_{\mathrm{var}}$ against $E$ and if $E_{\mathrm{res}}$ satisfies the inequality (\ref{eq:eres1}) we can confidently extrapolate to the point where the curve crosses the $E$-axis to find $E_{\mathrm{gs}}$. 

When we have several data points near $(E,\Delta_{\mathrm{var}})$ and use them for extrapolation, the slope is
\begin{equation}
    \frac{\partial E}{\partial \Delta_{\mathrm{var}}} = 1/(D_2/D_1 - 2E_{\mathrm{res}}),
\end{equation}
and the estimated energy via extrapolation is then
\begin{equation}
\begin{aligned}
    E_{\mathrm{extrp}} = &E - \Delta_{\mathrm{var}} \frac{\partial E}{\partial \Delta_{\mathrm{var}}}\\&=E-\frac{(D_2/D_1) E_{\mathrm{res}} - E_{\mathrm{res}}^2}{D_2/D_1 - 2E_{\mathrm{res}}}\\&=E_{\mathrm{gs}} + \frac{E^2_{\mathrm{res}}}{D_2/D_1 - 2E_{\mathrm{res}}}.
\end{aligned}
\end{equation}
In our numerical simulation, $E_{\mathrm{extrp}}$ is always slightly higher than $E_{\mathrm{gs}}$ since $D_2/D_1 > 2E_{\mathrm{res}}$ and the second term is positive.

If the quantum circuit is deep and noisy, we can approximately describe the effects of noise by a global depolarizing noise channel, $i.e.$, the final state is
\begin{equation}
    \rho_f = (1-\epsilon')|\psi_f\rangle\langle\psi_f| + \epsilon'\frac{I}{N},
\end{equation}
where $\epsilon'$ is the noise rate. (Note that $|\psi_f\rangle$ is usually not exactly the ideal state $|\psi_f\rangle$ we want in practical scenarios, there is a coherent mismatch~\cite{koczor2021exponential}.) Then, the variance is $\Delta_{\mathrm{var}}'=\operatorname{Tr}(\rho_f H^2)-\operatorname{Tr}^2(\rho_f H)$, the energy is $E' = \operatorname{Tr}(\rho_f H)$, and the residual energy is $E'_{\mathrm{res}} = E'-E_{\mathrm{gs}}$. Supposing $H_f$ is traceless, then together with equality (\ref{eq:var_e}), we obtain
\begin{equation}
    \Delta_{\mathrm{var}}' = \frac{D_2}{D_1}E'_{\mathrm{res}} - E'^2_{\mathrm{res}} + \epsilon'(\frac{D_2}{D_1}E_{\mathrm{gs}} + E_{\mathrm{gs}}^2 + \frac{\operatorname{Tr}(H_f^2)}{N}).
\end{equation}
We hope that $E'_{\mathrm{res}} \ll D_2/D_1$ and $\epsilon'$ is sufficiently small such that the first term $(D_2/D_1)E'_{\mathrm{res}}$ is dominant. In this case, the linear extrapolation still works. These conditions are usually satisfied in our experiments. However, note that there do exist cases where $(D_2/D_1)E'_{\mathrm{res}}$ is not dominant and variance extrapolation does not apply. For example, let us consider the case where the state only has non-zero overlaps with the ground state and the first excited state, $i.e.$, $c_1 = 1$, $c_j = 0$ for all $j > 1$. Then $D_2/D_1$ is the gap between the ground state and the first excited state which should be very small compared with $|E_{\text{gs}}|$ when the Hamiltonian is traceless. For realistic noise rate $\epsilon'$ of at least a few percent on current devices, $\epsilon'E^2_{\text{gs}}$ is non-negligible compared to $D_2/D_1E'_{\text{res}}$. 

When $\epsilon'$ becomes too large for variance extrapolation, we can use the virtual distillation technique~\cite{koczor2021exponential, PhysRevX.11.041036} to effectively suppress it to a smaller value. In ``virtual distillation", $\epsilon'$ decreases exponentially with increasing number of copies $M$. The required shot number, however, also increases exponentially with $M$.

Denoting $C' = (D_2/D_1)E_{\mathrm{gs}} + E_{\mathrm{gs}}^2 + \operatorname{Tr}(H^2)/N$, we find the estimated ground state energy via extrapolation to be
\begin{equation}
\begin{aligned}
    E'_{\mathrm{extrp}} = &E' - \Delta_{\mathrm{var}}' \frac{\partial E'}{\partial \Delta_{\mathrm{var}}'}\\&=E_{\mathrm{gs}} + \frac{E'^2_{\mathrm{res}} - \epsilon'C'}{D_2/D_1 - 2E_{\mathrm{res}}'}.
\end{aligned}
\end{equation}

\begin{figure}[h]
	\centering
	\includegraphics[width=6cm]{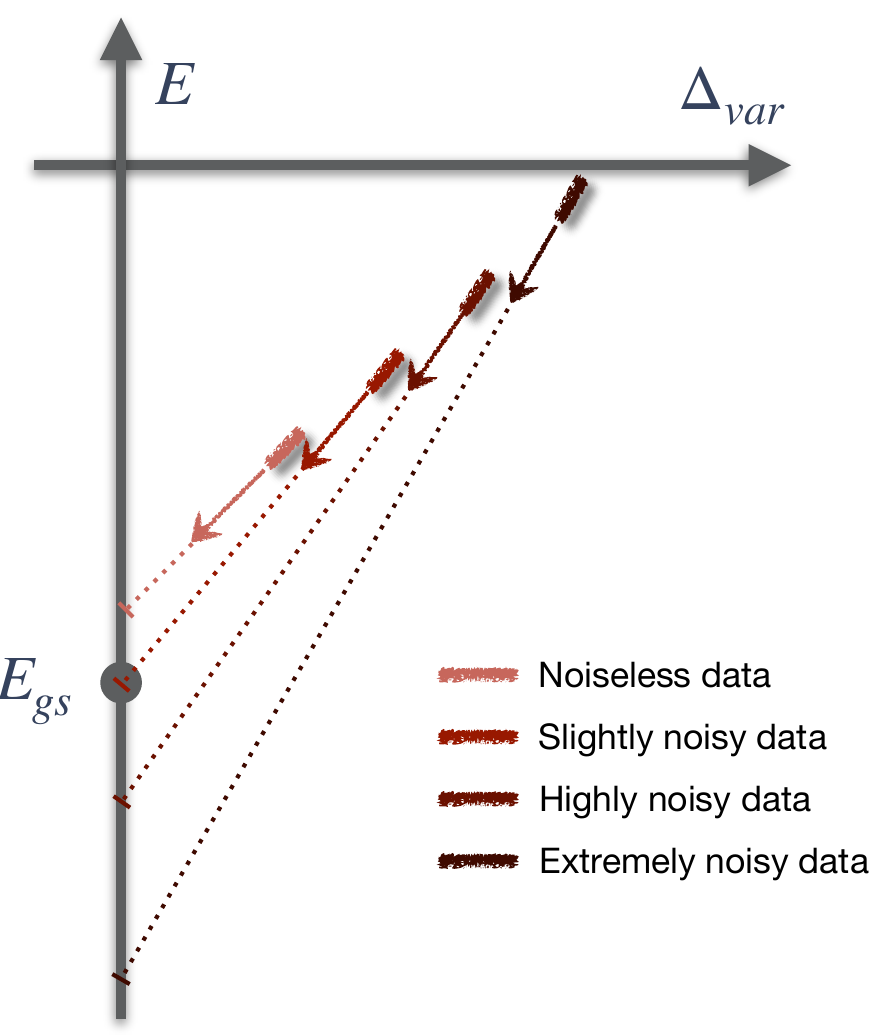}
	\caption{Schematic of the noise effect on variance extrapolation. Darker lines correspond to higher depolarizing noise rates. Arrows denote the direction of linear extrapolation. The most accurate estimate of the ground state energy is obtained in the slightly noisy case.}
	\label{fig:noise_effect}
\end{figure}

An illustration of variance extrapolation under different noise rates is given in Fig.~\ref{fig:noise_effect}. For a traceless Hamiltonian, global depolarizing noise increases the measured variance and damps the measured energy towards zero. With the increase of noise rate, the estimation error decreases first, then increases.

~\\

\bibliography{ref}

\end{document}